\documentclass[letterpaper,twocolumn,english,prl]{revtex4-1}
\usepackage[T1]{fontenc}
\usepackage[latin9]{inputenc}
\usepackage{babel}
\usepackage{mathtools}
\usepackage{amsmath}
\usepackage{amssymb}
\usepackage{graphicx}
\usepackage[unicode=true,
 bookmarks=true,bookmarksnumbered=false,bookmarksopen=false,
 breaklinks=false,pdfborder={0 0 1},backref=false,colorlinks=false]
 {hyperref}

\makeatletter

\pdfpageheight\paperheight
\pdfpagewidth\paperwidth

\usepackage{babel}
\usepackage{color}

\newcommand{\kT}{T}

\newcommand{\tf}{t_f}

\newcommand{\trajvec}{\boldsymbol{\varphi}}
\newcommand{\trajbackvec}{\tilde{\boldsymbol{\varphi}}}

\newcommand{\trajOnevec}{{\trajvec}^{(1)}}
\newcommand{\trajOnereversevec}{\tilde{\trajvec}^{(1)}}

\newcommand{\trajTwo}{\boldsymbol{\psi}}

\newcommand{\avec}{\mathbf{a}}
\newcommand{\avecunderline}{\underline{\avec}}

\newcommand{\Fvec}{\boldsymbol{F}}
\newcommand{\nablavec}{\boldsymbol{\nabla}}
\newcommand{\xvec}{\boldsymbol{x}}
\newcommand{\Xvec}{\boldsymbol{x}}
\newcommand{\trajTwovec}{\boldsymbol{\trajTwo}}

\newcommand{\timescale}{\tau}

\newcommand{\ti}{0}
\newcommand{\tfinal}{\tf}
\newcommand{\tinitial}{\ti}

\newcommand{\aexit}{\alpha_{R}}

\newcommand{\td}{\tau_D}

\newcommand{\RefillTime}{\Delta \mathcal{T}}

\newcommand{\xveci}{\xvec_{0}}
\newcommand{\xvecf}{\xvec_{f}}

\newcommand{\etavec}{\underline{\boldsymbol{\eta}}}

\makeatother

\begin{document}
\title{Measurement of irreversibility and entropy production via the tubular ensemble}
\author{Julian Kappler}
\email{jkappler@posteo.de}
\affiliation{Department of Applied Mathematics and Theoretical Physics, Centre for Mathematical Sciences, University of Cambridge, Wilberforce Road, Cambridge CB3 0WA, United Kingdom}
\author{Ronojoy Adhikari}
\affiliation{Department of Applied Mathematics and Theoretical Physics, Centre for Mathematical Sciences, University of Cambridge, Wilberforce Road, Cambridge CB3 0WA, United Kingdom}
\date{\today}
\begin{abstract}
The appealing theoretical measure of irreversibility in a stochastic process, as the ratio of the probabilities of a trajectory and its time reversal, cannot be accessed directly in experiment since the probability of a single trajectory is zero. We regularize this definition by considering, instead, the limiting ratio of probabilities for trajectories to remain in the tubular neighborhood of a smooth path and its time reversal. The resulting pathwise medium entropy production agrees with the formal expression from stochastic thermodynamics, and can be obtained from measurable tube probabilities. Estimating the latter from numerically sampled trajectories for Langevin dynamics yields excellent agreement with theory. 
By combining our measurement of pathwise entropy production with a Markov Chain Monte Carlo algorithm,
we infer the entropy-production distribution for a transition path ensemble directly from short 
recorded trajectories.
Our work enables the measurement of irreversibility along individual paths, and  path ensembles, 
in a model-free manner.
\end{abstract}
\maketitle

\section{Introduction}
Stochastic processes without memory have been used to describe the
dynamics of physical systems starting with the pioneering work of
Rayleigh, Einstein and Smoluchoswki \cite{chandrasekhar_stochastic_1943}.
The phenomenological observation that systems out of equilibrium display
irreversibility has prompted a search for theoretical measures that
enable its quantification. The first of these measures was provided
by Kolmogorov \cite{kolmogoroff_zur_1937,yaglom_statistical_1949}
by considering the joint distribution of pairs of points along a stochastic
trajectory and its time reversal. This characterisation was refined
by Ikeda and Watanabe \cite{ikeda_stochastic_1989} by considering
the probabilities for stochastic trajectories to remain in the tubular
neighborhood of a smooth path and its time reversal. This line of
thought reached its culmination in the elementary definition of irreversibility
as the ratio of probabilities for a trajectory and its reverse in
the work of Maes and Neto\v{c}ny \cite{maes_time-reversal_2003} and Seifert \cite{seifert_entropy_2005}.
This provides the clearest derivation of the plethora of results know
as fluctuation theorems \cite{bochkov_general_1977,jarzynski_nonequilibrium_1997,kurchan_fluctuation_1998,crooks_entropy_1999,dalibard_origin_2004,chernyak_path-integral_2006,seifert_stochastic_2012},
yields a definition of the medium entropy production as the logarithm
of the ratio of the probability of forward and backward paths \cite{seifert_entropy_2005},
and has engendered the thriving field of stochastic thermodynamics
\cite{sekimoto_stochastic_2010,seifert_stochastic_2012,seifert_stochastic_2019}. 

Despite the theoretical importance of the elementary definition of
irreversibility, measurements, in both experiment and simulation,
have focussed on ensembles of trajectories \cite{luchinsky_irreversibility_1997,luchinsky_analogue_1998,otsubo_estimating_2020,manikandan_inferring_2020}
or systems with discrete state space \cite{tietz_measurement_2006},
and the medium entropy production along a single continuous trajectory
has not yet been measured directly. This is because the probability
of a trajectory (and of its reversal) is, strictly speaking, zero
and it is therefore not obvious how to infer the ratio of probabilities 
for a pair of forward and reverse path.
Hence, while theoretical expressions for this ratio can be evaluated on
observed trajectories, the result cannot be tested without an independent, 
model-free method of inferring pathwise irreversibility.

In this work, we provide a resolution to this impasse by considering,
instead of a single trajectory, the probability of an ensemble of
trajectories to remain within the tubular neighborhood of a smooth
path \cite{kappler_stochastic_2020,gladrow_direct_2020}. We define
the logarithm of the probability ratio for forward and backward tubes, as
the tube radius goes to zero, as a measure of irreversibility. We
show that this coincides with the stochastic thermodynamic expression
for the medium entropy production when the latter is restricted to
smooth paths. Since the probability to remain within a finite-radius tube
can be measured 
directly \cite{gladrow_direct_2020}, we obtain
the medium entropy production by extrapolating
 ratios of measured finite-radius tube probabilities to the limit of vanishing radius.
This requires no knowledge of the underlying process (other
than that it is memoryless) and our method, then, yields a model-free
route to obtaining the entropy production along individual paths.
\textcolor{black}{This establishes
a protocol for directly measuring irreversibility along individual
pathways, and allows us to investigate this phenomenon, experimentally
or numerically, in a manner that is far more refined than full
 ensemble averages.}
\textcolor{black}{We validate our method in an explicit numerical example.} 
For two-dimensional Langevin dynamics with a non-equilibrium force,
 we directly infer the medium entropy production
along individual paths \textcolor{black}{from simulated trajectories without using any 
knowledge about the underlying dynamics beyond Markovianity}, 
and find excellent agreement with the theoretical
expectation \cite{seifert_entropy_2005}.
Furthermore, 
by combining the direct measurements of relative path probabilities \cite{gladrow_direct_2020},
our approach to the single-trajectory entropy production,
and a Metropolis-Hastings
Markov Chain Monte Carlo (MCMC) algorithm \cite{thijssen_computational_2007}, we
infer the distribution of the entropy production for a transition path ensemble
directly from measured sojourn probabilities.

\textcolor{black}{
The remainder of this paper is organized as follows.
In Sect.~\ref{sec:tubular_entropy_section}, we define the medium entropy production
as limiting ratio of tube probabilities, and discuss analytically the special case for
overdamped It\^{o}-Langevin dynamics.
We subsequently explain how we infer finite-radius tube probabilities from
recorded time series in practice \cite{gladrow_direct_2020}.
In Sect.~\ref{eq:two_dimensional_example} we then consider
 a two-dimensional example system.
Using a dataset generated via numerical simulations, we first 
 measure the entropy production along individual paths
and compare the results to analytical predictions.
We then go on to infer the entropy production distribution for a transition path ensemble,
using only measured tube probabilities, 
and compare the results to an independently generated
transition path ensemble based on direct Langevin simulations.
We analyze these transition path ensembles further, by considering both paths with very small
entropy production, and paths with very large entropy production.
In Sect.~\ref{sec:discussion} we close by summarizing our results
and discussing their further implications.
}

\section{Irreversibility via asymptotic tube probabilities} 
\label{sec:tubular_entropy_section}

\subsection{Entropy production as asymptotic ratio of tube probabilities}
For a
smooth reference path $\trajvec_{t},t\in[\ti,\tf]$, we define the
sojourn probability that a stochastic trajectory $\Xvec_{t}$ remains
within a tube of radius $R$ around $\trajvec$ as $P_{R}^{\trajvec}(t)\equiv P(\,||\Xvec_{s}-\trajvec_{s}||<R~\forall s\in[\ti,t]\,)$,
where $||\boldsymbol{{v}}||=(v_{1}^{2}+v_{2}^{2}+...+v_{N}^{2})^{1/2}$
denotes the standard Euclidean norm in $\mathbb{R}^{N}$, and where
we suppress the dependence on the initial condition of the trajectory
inside the tube \cite{kappler_stochastic_2020}. Combining the approach
to irreversibility via tubes \cite{ikeda_stochastic_1989} with the
single-trajectory medium entropy production \cite{seifert_entropy_2005,seifert_stochastic_2012},
we define the medium entropy change along $\trajvec$ in terms of asymptotic tube
probabilities as 
\begin{equation}
\Delta s_{\mathrm{m}}[\trajvec]\equiv\lim_{R\rightarrow0}\ln\frac{P_{R}^{\trajvec}(\tf)}{P_{R}^{\trajbackvec}(\tf)},\label{eq:sm_sojourn}
\end{equation}
with $\trajbackvec_{t}\equiv\trajvec_{\tf-t}$ the time-reverse of
the path $\trajvec$.
\textcolor{black}{In our definition Eq.~\eqref{eq:sm_sojourn} we 
assume that temperature $T$ is measured
in units of energy, so that entropy is dimensionless \cite{ben-naim_farewell_2008}.}
 Equation \eqref{eq:sm_sojourn}
relates the medium entropy production along a single path to
observable sojourn probabilities.
For finite radius $R$, the ratio of sojourn probabilities 
for forward
and backwards path
can be measured  without fitting a model
to the data, by simply  counting how many sample trajectories leave
the tube along forward and backward path, respectively \cite{kappler_stochastic_2020,gladrow_direct_2020}.
According to Eq.~\eqref{eq:sm_sojourn},
performing this measurement for several finite values of
$R$, and extrapolating the 
resulting log-ratios
 to $R\rightarrow0$, the medium
entropy production $\Delta s_{\mathrm{m}}$ is obtained.

The decay of the sojourn probability 
is described by $\aexit^{\trajvec}(t)$, the instantaneous exit rate
with which stochastic trajectories first leave the tube, as \cite{kappler_stochastic_2020}
\begin{equation}
\label{eq:exit_def}
\aexit^{\trajvec}(t)\equiv - \frac{ 
					(\partial_{t}{P}_{R}^{\trajvec})(t)
					}{
					P_{R}^{\trajvec}(t)}\,,
\end{equation}
Differentiating Eq.~\eqref{eq:sm_sojourn} with respect to $t_f$,
 substituting $t_f$ with $t$,
 and eliminating sojourn probabilities in favor of exit rates,
we obtain
\begin{align}
\frac{d}{dt}\Delta s_{\mathrm{m}}[\trajvec] & =-
\lim_{R\rightarrow0}
\Delta\aexit^{\trajvec}(t)\,,
\label{eq:definition_path_entropy_in_terms_of_exit_rate} 
\end{align}
where
\begin{align}
\Delta\aexit^{\trajvec}(t) & \equiv\aexit^{\trajvec}(t)-\aexit^{\trajbackvec}(\tf-t).\label{eq:finite_radius_rate_difference}
\end{align}
Equation \eqref{eq:definition_path_entropy_in_terms_of_exit_rate} relates the 
 change in medium entropy production along a single path to the difference of
 instantaneous tubular exit rates around forward and backward versions of the path.

\subsection{Analytical results for Langevin dynamics}
While
Eqs.\,(\ref{eq:sm_sojourn}-\ref{eq:finite_radius_rate_difference})
do not assume a model for the stochastic evolution of $\Xvec_{t}$,
for a given model the exit rate can be calculated analytically. We
now consider the overdamped It\^{o}-Langevin equation for an $N$-dimensional coordinate
$\Xvec_{t}\equiv(x_{1}(t),x_{2}(t),...,x_{N}(t))$, given 
by 
\begin{equation}
\mathrm{d}\Xvec_{t}=\mu\boldsymbol{F}(\Xvec_{t})\,\mathrm{{d}}t+\sqrt{2\mu \kT}\,\mathrm{{d}}\boldsymbol{W}_{t}\,,\label{eq:Langevin}
\end{equation}
where $\mu=D/\kT$ is the mobility with $D$ the diffusion coefficient and $T$ the absolute temperature
measured in units of energy,
$\Fvec$ is a deterministic force, and $\mathrm{d}\boldsymbol{W}_{t}$ denotes the increment
of the
Wiener process. While we here only consider forces
that do not depend on time explicitly, our approach remains valid
for time-dependent forces as long as for time-reversed paths the explicit
time-dependence of the force is also reversed \cite{seifert_stochastic_2012}.
For Eq.~\eqref{eq:Langevin}, the leading-order expansion of $\aexit^{\trajvec}(t)$
in the tube radius $R$ is \cite{ito_probabilistic_1978,fujita_onsager-machlup_1982,kappler_stochastic_2020}
\begin{equation}
\aexit^{\trajvec}(t)=\frac{C_{N}}{R^{2}}+\mathcal{L}^{\trajvec}(t)+\mathcal{{O}}(R^{2}),\label{eq:aexit_series}
\end{equation}
where $C_{N}$ is a constant which only depends on the dimension $N$,
and the Onsager-Machlup (OM) Lagrangian $\mathcal{L}^{\trajvec}$
is given by \cite{onsager_fluctuations_1953,durr_onsager-machlup_1978,ito_probabilistic_1978,williams_probability_1981,fujita_onsager-machlup_1982,kappler_stochastic_2020}
\begin{equation}
\mathcal{L}^{\trajvec}=\frac{1}{4D}\left(\dot{\trajvec}-\mu\Fvec(\trajvec)\right)^{2}+\frac{1}{2}\text{div}\left(\,\mu\Fvec(\trajvec)\right).\label{eq:OnsagerMachlup}
\end{equation}
Substituting Eq.\,\eqref{eq:OnsagerMachlup} into the difference
of exit rates Eq.\,\eqref{eq:finite_radius_rate_difference} for forward- and reverse
path, the relation
\begin{equation}
\lim_{R\rightarrow0}\Delta\aexit^{\trajvec}(t)=-\frac{1}{T}\Fvec(\trajvec_{t})\cdot\dot{\trajvec}_{t}\label{eq:rate_and_work_rate}
\end{equation}
between the limit of exit-rate difference and work rate along $\trajvec$
follows. In turn substituting this into Eq.~\eqref{eq:definition_path_entropy_in_terms_of_exit_rate},
and integrating with respect to time,
yields the familiar formula \cite{seifert_entropy_2005,seifert_stochastic_2012}
\begin{equation}
\Delta s_{\mathrm{m}}[\trajvec]=\frac{1}{\kT}\int_{\ti}^{\tf}\Fvec(\trajvec_{t})\cdot\dot{\trajvec}_{t}\,\mathrm{d}t,\label{eq:theory_entropy_production_general}
\end{equation}
which relates the medium entropy production and the work performed
along $\trajvec$.

\subsection{Measuring the exit rate from sample trajectories}
\label{sec:algorithm_main}

\begin{figure}[ht!]
\centering \includegraphics[width=1\columnwidth]{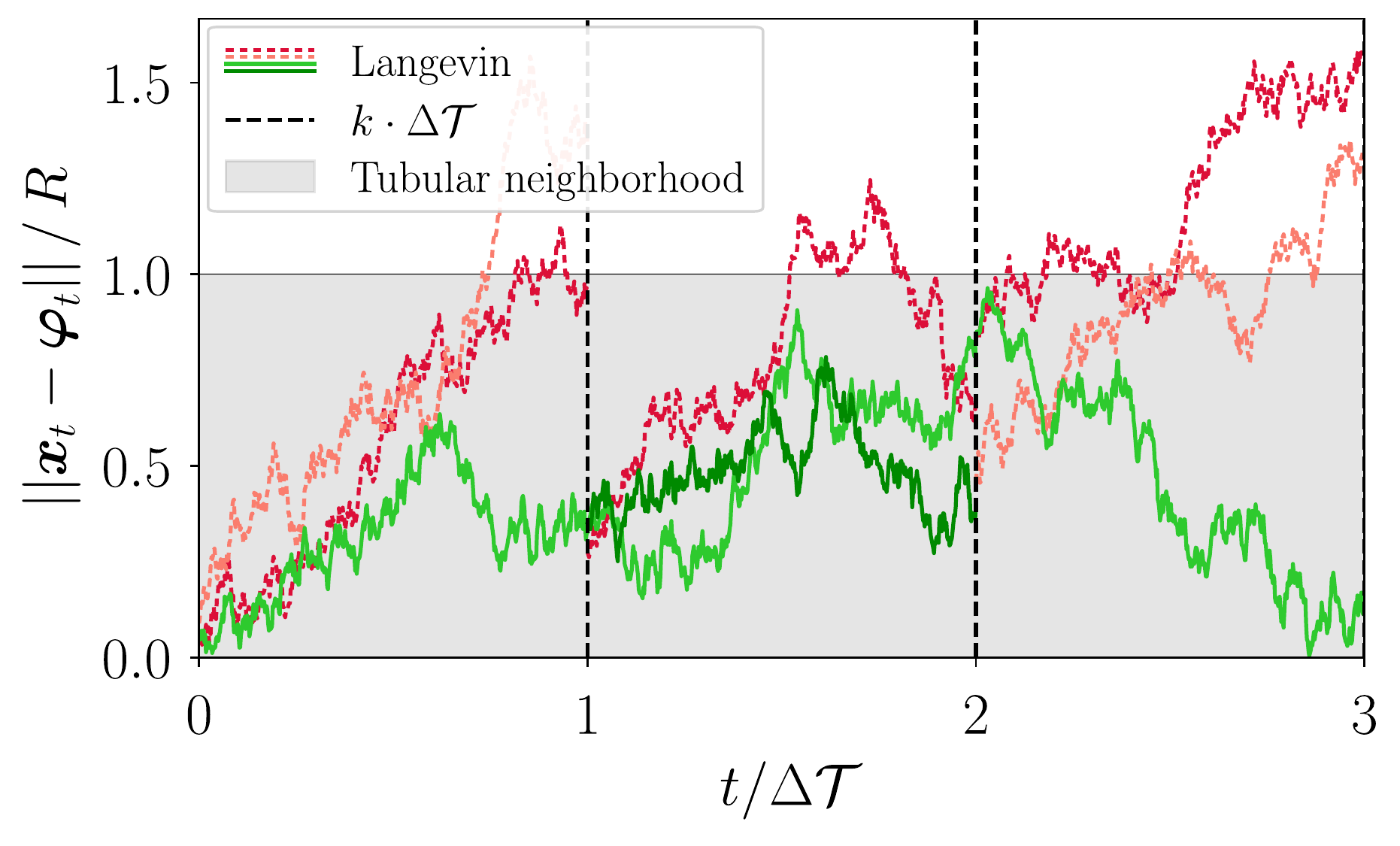}
\caption{ \label{fig:cloning_algorithm} 
\textcolor{black}{The gray shaded area denotes a tube
of radius $R$ around a reference path $\trajvec$. 
From a given set of short sample trajectories, 
we randomly draw
$M_0=3$  samples of duration $\RefillTime$
(vertical dashed lines), all of which start close to $\trajvec_{\ti}$.
Trajectories which leave the tube (dotted red lines) are discarded,
the final positions of the trajectories that stay (solid green lines)
are collected.
We again draw $M_1=3$ sample trajectories, with initial conditions approximately
distributed as the previous final positions. 
We repeat the process of drawing sample trajectories and tracking whether 
or not they leave the tube,
until we obtain the sojourn probability up to the desired final time.
For each interval $[l \RefillTime, (l+1) \RefillTime]$, 
the sojourn probability is estimated as the fraction of 
 trajectories that leave the tube during the time $\RefillTime$, 
 divided by 
 the initial number of trajectories $M_l$.
We choose the small value $M_l=3$ here for
illustration; to calculate exit rates from simulations we use values
of the order $10^{5}$, see App.~\ref{sec:cloning_algorithm}. 
For the data shown here we use the reference
path $\trajvec\equiv\trajOnevec=L(t/\tfinal,t/\tfinal)$, and 
the force Eq.~\eqref{eq:shear_force_sum}
with $\theta=1$, $L F_{0}/T=5$. Furthermore,
we use $R=0.3L$, $\RefillTime=0.01\,\tau$, and
$D = L^2/\tau$.
}
}
\end{figure}

 To infer the entropy
production from finite-radius exit rates, the log-ratio on the
 right-hand side of Eq.~\eqref{eq:sm_sojourn}
needs to be measured for small but finite radius $R$. In practice
it can be difficult to acquire sufficient data for this measurement,
because the number of trajectories which remain inside the tube decreases
exponentially with time $t$. 
To overcome this problem, we employ a cloning algorithm, 
which
\textcolor{black}{is
illustrated in Fig.~\ref{fig:cloning_algorithm} and}
\textcolor{black}{has previously been used
to infer finite-radius exit rates from one-dimensional experimental time series
\cite{gladrow_direct_2020}.
We here present a short summary of the algorithm, and give more details in
App.~\ref{sec:cloning_algorithm}.
}

The cloning algorithm assumes that the underlying stochastic dynamics
 is  Markovian, 
 and that an ensemble of 
 recorded short trajectories with initial conditions throughout the domain of interest are available;
these can originate either from measurements \cite{gladrow_direct_2020}, or, as in this work, from simulations.

\textcolor{black}{
 For a given reference
path $\trajvec$ and tube radius $R$,
we initialize the algorithm by drawing from the ensemble of recorded trajectories 
 a large number $M_0$ of sample trajectories,
with initial conditions close to $\trajvec_{\tinitial}$.
We then follow those sample trajectories for a short time $\RefillTime$, 
and discard each trajectory once it leaves a moving ball of radius $R$
and instantaneous center $\trajvec_t$ for the first time. 
We then estimate the sojourn probability $P_{R}^{\trajvec}(t)$ at time $t\in [\tinitial,\RefillTime]$
 by the fraction of sample trajectories that have never
 left the tube until the time $t$.
}

\textcolor{black}{
To iteratively 
obtain the sojourn probability also for any subsequent time interval $[l \RefillTime, (l+1) \RefillTime]$,
we in each iteration step draw $M_l$ sample trajectories with initial conditions inside a
ball of radius $R$ and with center $\trajvec(l \RefillTime)$.
For the initial distribution within the ball, we at each iteration use the final spatial distribution
of those trajectories that have never left the tube in the previous iteration.
We repeat this iteration step until $l \RefillTime = \tfinal$.
}

\textcolor{black}{
By periodically drawing new samples after a short time $\RefillTime$, we overcome
the exponential decay of the trajectories that have never left the tube.
For a given $\RefillTime$, we choose the number of trajectories
$M_l$  dynamically based on the expected decay of the sojourn probability during
the time interval $[l \RefillTime, (l+1) \RefillTime]$, 
as we explain in detail in App.~\ref{sec:cloning_algorithm}.
In practice, one wants to choose the time $\RefillTime$ so as to
 balance the exponential decay of the sojourn probability 
with the cost of re-drawing sample trajectories: 
If $\RefillTime$ is too large, a large number of sample trajectories $M_l$ 
is required
to reliably estimate the sojourn probability for the whole time interval $[l \RefillTime, (l+1) \RefillTime]$.
On the other hand, if $\RefillTime$ is too small, new trajectory samples have to be drawn 
very frequently.
}

\section{Two-dimensional non-equilibrium example} 
\label{eq:two_dimensional_example}

\begin{figure*}[ht!]
\centering \includegraphics[width=\textwidth]{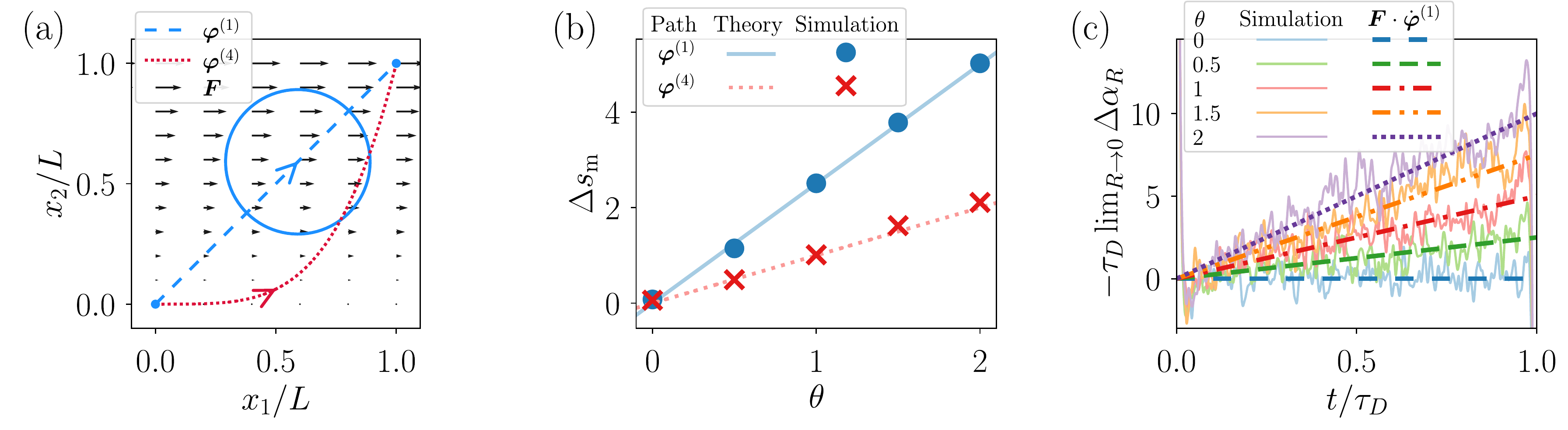}
\caption{\label{fig:shear_example} \textit{(a) } The shear force Eq.~\eqref{eq:shear_force_sum}
is shown as black quiver plot. The path 
defined in
Eq.~\eqref{eq:shear_path0}
 is shown for $n=1$ (dashed blue line)
and $n=4$ (dotted red line), with arrows indicating the forward direction.
For $\trajOnevec$ we include a snapshot of the instantaneous tube
of radius $R/L=0.3$ around the path (blue circle). 
\textit{(b)} The
colored lines denote the theoretical entropy production, Eq.~\eqref{eq:shear_sm_theory_0},
as a function of $\theta$ for $n=1$ (solid blue line) and $n=4$ (dotted red line). 
The
colored symbols are obtained by
 extrapolating measured finite-radius log-ratios of sojourn probabilities to $R=0$,
to evaluate Eq.~\eqref{eq:sm_sojourn}.
\textit{(c)} The colosolid red lines denote the
negative of the left-hand side of Eq.~\eqref{eq:rate_and_work_rate},
obtained by extrapolating measurements of Eq.~\eqref{eq:finite_radius_rate_difference}, for the reference
path $\trajvec^{(1)}$,
to $R = 0$.
 The colored broken lines denote the corresponding
theoretical predictions given by the negative of the right-hand side of Eq.~\eqref{eq:rate_and_work_rate},
calculated using the force Eq.~\eqref{eq:shear_force_sum} with $LF_{0}/T=5$.
Numerical data in (c) is smoothed using a Hann window of width $0.015\,\td$.  
}
\end{figure*}

\subsection{Model system}
For a length
scale $L$ and a time scale $\timescale$, we consider
Eq.~\eqref{eq:Langevin} for dimension $N=2$ with diffusivity $D=L^{2}/\timescale$,
so that $\tau_{D}\equiv L^{2}/D=\timescale$. We consider a shear
force 
\begin{equation}
\Fvec(\xvec)=
\frac{\theta F_{0}}{L}\begin{pmatrix}x_{2}\\
0
\end{pmatrix},\label{eq:shear_force_sum}
\end{equation}
where we fix $L F_{0}/T=5$, so that the dimensionless
parameter $\theta\in\mathbb{R}$ controls the force amplitude.
Equation \eqref{eq:shear_force_sum} does not admit a potential, and
is illustrated as a quiver plot in Fig.~\ref{fig:shear_example}
(a).
For each of the values $\theta=0,0.5,1,1.5,2$,
we generate an independent set of short sample Langevin trajectories with random initial conditions,
as described in detail in App.~\ref{sec:sample_data_preparation}.

\subsection{Entropy production along individual paths}

 We consider a family of paths 
\begin{align}
\trajvec_{t}^{(n)} & =L\begin{pmatrix}t/\tfinal\\[1.2ex]
(t/\tfinal)^{n}
\end{pmatrix},\label{eq:shear_path0}
\end{align}
where $t\in[\ti,\tf]\equiv[0,\timescale]$\textcolor{black}{, 
the length scale $L$ multiplies both vector components,}
 and $n\in\mathbb{N}$ enumerates
the paths. For any $n$, the path starts at $\xveci=(0,0)$ and ends
at $\xvecf=(L,L)$, example paths for $n=1$ and $n=4$ are shown in
Fig.~\ref{fig:shear_example} (a). 
For this family of paths
and the force Eq.~\eqref{eq:shear_force_sum}, the analytical medium
entropy production Eq.~\eqref{eq:theory_entropy_production_general}
 evaluates to 
\begin{equation}
\Delta s_{\mathrm{m}}[\trajvec^{(n)}]=
\frac{L F_{0}}{T(n+1)}\theta.
\label{eq:shear_sm_theory_0}
\end{equation}

We now consider the case $n=1$. 
We use the cloning algorithm
described in Sect.~\ref{sec:algorithm_main} and App.~\ref{sec:cloning_algorithm}
 with $\RefillTime = 0.01\td$ to measure
 the finite-radius sojourn probability for both $\trajOnevec$ and
its time-reverse $\trajOnereversevec$ for radius $R/L=0.3,0.4,0.5,0.6,0.7$.
 We fit a quadratic function $f(R)=a+R^{2}b$
to the resulting measured log-ratios $\ln{P_{R}^{\trajvec}(\tf)}/{P_{R}^{\trajbackvec}(\tf)}$, 
and extrapolate to zero as
$\lim_{R\rightarrow0}\ln{P_{R}^{\trajvec}(\tf)}/{P_{R}^{\trajbackvec}(\tf)} = a$,
where here $\trajvec \equiv \trajOnevec$.
The quadratic form of the fit function $f(R)$ is motivated by Eq.~\eqref{eq:aexit_series}, 
according to which no terms linear in $R$ appear in the exit rate, and hence in the log-ratio of
 sojourn probabilities.
In Fig.~\ref{fig:shear_example} (b), the  extrapolated measured
log-ratio is compared to the corresponding
analytical expectation, given by the right-hand side of Eq.~\eqref{eq:shear_sm_theory_0}.
We observe that the measurement agrees very well with the theoretical
prediction, which shows that Eq.~\eqref{eq:sm_sojourn} 
can be used to infer the irreversibility
along individual paths directly from data.
This is further confirmed by repeating the analysis protocol for a second path, 
where $n=4$. Figure \ref{fig:shear_example} (b) shows that also here,
 the extrapolated log-ratio obtained
from direct measurement agrees very well with the theoretical prediction.

To see how the medium entropy production is partitioned along the path, 
we evaluate the instantaneous exit rate for forward- and backward path for the $n=1$ path.
For this we discretize
 Eq.~\eqref{eq:exit_def}
 using central finite differences, and evaluate the expression on
 the measured sojourn probability for $R/L = 0.3$, $0.4$, $0.5$, $0.6$, $0.7$.
To extrapolate the resulting finite-radius measurements of 
Eq.~\eqref{eq:finite_radius_rate_difference}
to the limit $R \rightarrow 0$, we at each recorded time $t$ fit a quadratic function 
$f(t)=a(t)+R^{2}b(t)$ to the measured 
exit-rate difference $\Delta\aexit^{\trajvec}(t)$.
From this fit, we obtain the extrapolated exit-rate difference at time $t$ as 
$\lim_{R\rightarrow0}\Delta\aexit^{\trajvec}(t) = a(t)$ \cite{gladrow_direct_2020}.
In Fig.~\ref{fig:shear_example} (c)
we compare the resulting extrapolated exit-rate differences 
to the theoretical expectation 
Eq.~\eqref{eq:rate_and_work_rate}.
While overall the agreement between measurement and
theory is very good, there are deviations both in the
beginning, $t\lesssim0.05\,\td$, and at the end of the trajectory,
$t\gtrsim0.95\,\td$.
This is
because in our cloning algorithm all initially sampled trajectories
 start close to the center of the tube, so that at the beginning/end
we observe the initial relaxation of this initial condition for the
forward/reverse path \cite{kappler_stochastic_2020}.
The agreement in Fig.~\ref{fig:shear_example} (b) shows that these transient effects 
 are not important for the integrated change in medium entropy production, i.e.~for
 $\Delta s_{\mathrm{m}}$.

In App.~\ref{app:double_well} we consider another
two-dimensional example system, comprised of a circular double-well
potential superimposed with a circular non-equilibrium force; the example
again confirms the validity and practical applicability of Eqs.~\eqref{eq:sm_sojourn}, \eqref{eq:definition_path_entropy_in_terms_of_exit_rate}.

\subsection{Medium entropy production for transition-path ensemble}
\label{sec:medium_entropy_production}
We now infer the entropy-production distribution for an ensemble of transition paths,
using only measured sojourn probabilities.
For this, we use the dataset of Langevin time series
corresponding to the force Eq.~\eqref{eq:shear_force_sum},
with $L  F_0/T = 5$, $\theta = 1$,
and $D = L^2/\tau$.
We consider continuous paths which start at $\xveci =(0,0)$ at time $t = \tinitial$, and end
at $\xvecf=(L,L)$ at time $t_f = \tau_D$.
We approximate the infinite-dimensional
space of all such paths by a path-space of dimension $d = N M$, 
\textcolor{black}{where for our two-dimensional system $N=2$.
Our finite-dimensional approximation of path space is  parametrized by a set of $M$
$N$-dimensional expansion coefficients}
$\underline{\avec} \equiv (\avec_1,..., \avec_M) \in \mathbb{R}^{N \times M}$.
For any $\underline{\avec}$, the corresponding path is then given by
\begin{equation}
\label{eq:path_parametrization}
\trajvec_t(\underline{\avec}) = \xvecf \frac{ t}{\tau_D} +  \sum_{k = 1}^{M} \frac{\avec_k}{k} \sin\left( k \pi  \frac{ t }{\tau_D}\right).
\end{equation}
We use $M = 15$, so that \textcolor{black}{for our $N=2$ dimensional system we have}  $d =30$,
and run a Metropolis-Hastings Markov Chain Monte Carlo (MCMC) algorithm \cite{thijssen_computational_2007}
 on the space $\mathbb{R}^d$,
to infer the distribution of $\Delta s_{\mathrm{m}}$;
we explain the algorithm in detail in App.~\ref{sec:MCMC}.
Crucially, the algorithm only uses measured sojourn probabilities.
First, to generate an ensemble of transition paths, 
ratios of path probabilities  need to be evaluated;
for this we use extrapolated log-ratios of 
measured finite-radius sojourn probabilities \cite{gladrow_direct_2020}.
Second, we obtain the medium entropy production along each 
 path from measured finite-radius
sojourn probabilities via
Eq.~\eqref{eq:sm_sojourn}.
Using the sojourn-probability MCMC algorithm, we generate
a set of
$57448 \approx 5.7 \times 10^{4}$
transition paths, 
and accompanying values for $\Delta s_{\mathrm{m}}$.

For comparison, we additionally generate an independent ensemble of transition paths.
Using the Euler-Maruyama integration scheme
with timestep $\Delta t/\tau_D = 10^{-4}$,
we run a large number of numerical simulations of Eq.~\eqref{eq:Langevin},
each of duration $\td$ and with initial condition $\xveci = (0,0)$.
We retain only those trajectories that at the final time are
 within the rectangle $[0.9 L,1.1L] \times [0.9 L, 1.1L]$ around  $\xvecf = (L,L)$.
Using this protocol, we create an ensemble of 
$3316727 \approx 3.3 \times 10^{6}$
Langevin transition paths. For each trajectory, we evaluate
 Eq.~\eqref{eq:theory_entropy_production_general}
to obtain the corresponding analytical prediction for $\Delta s_{\mathrm{m}}$ \cite{bo_functionals_2019}.

\begin{figure}[ht!]
\centering \includegraphics[width=0.9\columnwidth]{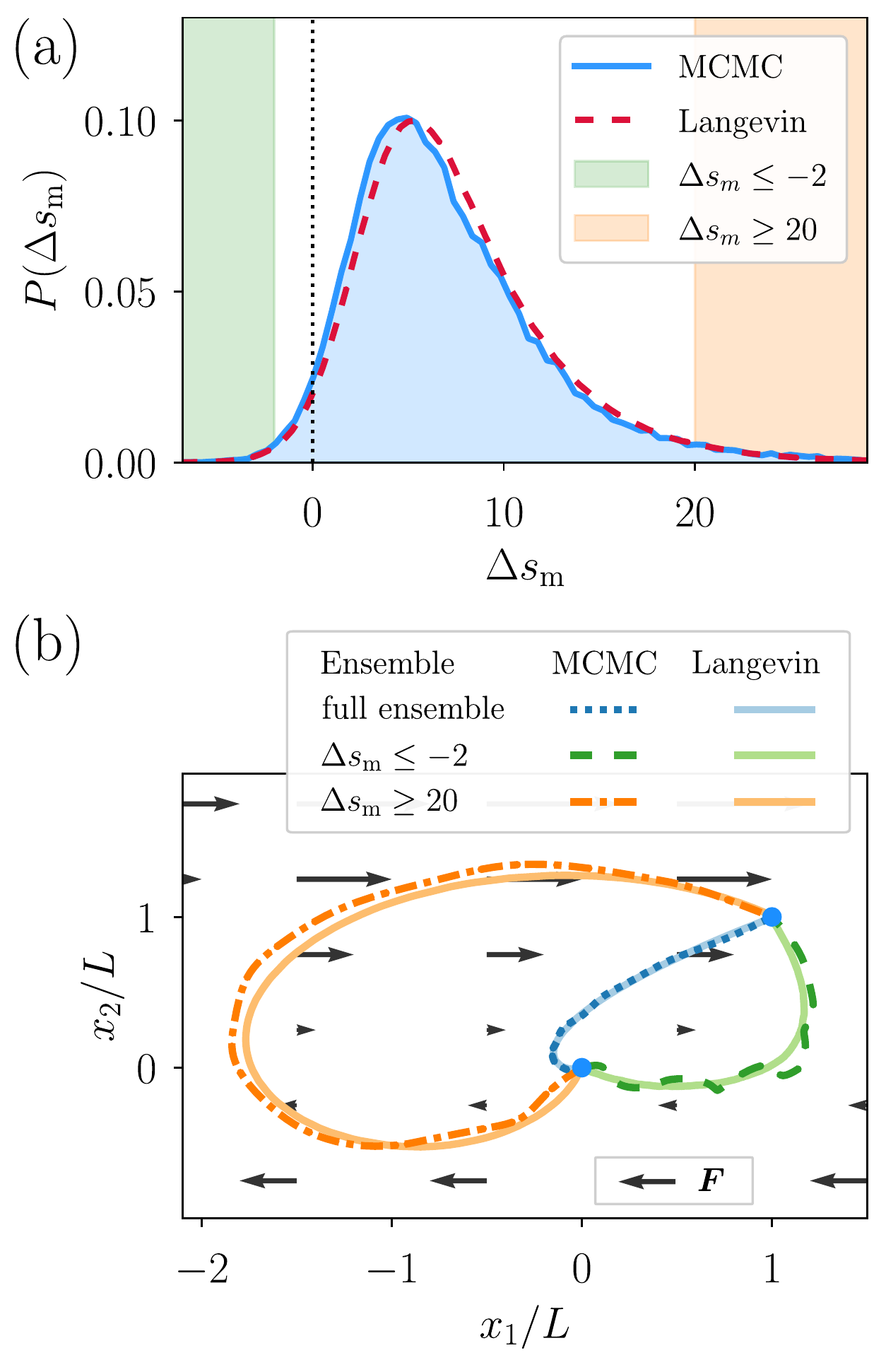}
\caption{
 \label{fig:MCMC} 
 \textit{(a)}
Distribution of $\Delta s_{\mathrm{m}}$ for the ensemble of transition paths
that start at $\xveci = (0,0)$ and are at $\xvecf = (L,L)$ 
a duration $\td$ later.
  The solid blue curve
  is  obtained from approximately $5.7 \times 10^4$ samples generated via
a MCMC algorithm, which only uses measured sojourn probabilities.
Full details on the algorithm are given in App.~\ref{sec:MCMC}.
The dashed red line is obtained 
from evaluating 
 Eq.~\eqref{eq:theory_entropy_production_general} on 
 an independently generated ensemble of transition paths, as
 explained in Sect.~\ref{sec:medium_entropy_production}.
The vertical dotted line denotes $\Delta s_{\mathrm{m}} = 0$,
the colored shaded regions depict 
the ranges $\Delta s_{\mathrm{m}} < -2$ (green),  and
 $\Delta s_{\mathrm{m}} \geq 20$ (orange).
\textit{(b)}
The solid lines depict the mean transitions paths inferred from
 the MCMC algorithm, the broken lines correspond to the mean 
 transition paths obtained from direct Langevin simulations.
While the dotted and solid blue lines correspond to the mean over the whole ensemble,
the green/orange lines depict the mean over all sample paths with 
$\Delta s_{\mathrm{m}} \geq 20$ (dashed-dotted and solid orange lines) 
and  $\Delta s_{\mathrm{m}} \leq -2$ (dashed and solid green lines).
For all data shown, we fix a length scale $L$ and a time scale $\tau$, and
 consider the force Eq.~\eqref{eq:shear_force_sum},
with $\theta=1$ and $LF_{0}/T=5$,
as well as $D = L^2/\tau$.
}
\end{figure}

In Fig.~\ref{fig:MCMC} (a) we show the resulting distribution for $\Delta s_{\mathrm{m}}$,
obtained from both the sojourn-probability MCMC and the 
 direct Langevin simulations. 
Our MCMC algorithm, which only uses measured sojourn probabilities
inferred from short recorded time series, reproduces 
the distribution of the medium entropy production for the ensemble very well;
the slight deviations between the two datasets can be explained by 
our low-dimensional approximation Eq.~\eqref{eq:path_parametrization} of path-space, 
as we discuss in App.~\ref{sec:MCMC_further_analysis}.

Both the sojourn-probability MCMC
and the direct Langevin simulations yield an ensemble of
 transition paths with accompanying entropy production.
We now  analyze this ensemble further, to gain insight into the mean behavior
of small- and large-entropy transition paths.

We first calculate the mean transition path from both ensembles, by
at each time $t$ averaging over the current positions of all trajectories in the
respective ensemble.
In Fig.~\ref{fig:MCMC} (b) we show the resulting mean transition paths
as dotted and solid blue lines,
which agree very well.

We next consider the subensemble of paths
with small entropy production $\Delta s_{\mathrm{m}} \leq -2$,
as indicated by the green shaded region on the left side of Fig.~\ref{fig:MCMC} (a). 
Out of our dataset of 57448 MCMC sample paths, 294 fulfill $\Delta s_{\mathrm{m}} \leq -2$,
so that we estimate the probability to observe any such small-entropy path as
 $P(\Delta s_{\mathrm{m}} \leq -2) \approx 294/57448 \approx 5.1\times 10^{-3}$.
This is close to the corresponding value $P(\Delta s_{\mathrm{m}} \leq -2) \approx 15697/3316727
\approx 4.7\times 10^{-3}$ estimated from the direct Langevin simulations.
We calculate the mean small-entropy path by at any time $t$ averaging over all
the positions of all paths with $\Delta s_{\mathrm{m}} \leq -2$,
and show the corresponding results 
in Fig.~\ref{fig:MCMC} (b) as dashed and solid green  curves.
Again, the MCMC result agrees well with the mean small-entropy path obtained from direct Langevin simulations;
we rationalize the oscillatory behavior of the MCMC path by the small number of 
samples.
From Fig.~\ref{fig:MCMC} (b) we observe that small-entropy paths on
 average move to $x_1 \approx L$
with very small negative value of $x_2$, and move up to $\xvecf$ slightly against the 
shear force.
This motion against the deterministic force is how these trajectories perform work, i.e.~how they obtain
 a negative medium entropy production.

We finally consider 
paths with large medium entropy production $\Delta s_{\mathrm{m}} \geq 20$,
as indicated by the orange shaded region on the right side of Fig.~\ref{fig:MCMC} (a).
In our MCMC ensemble, there are 1335 MCMC sample paths with $\Delta s_{\mathrm{m}} \geq 20$,
so that we estimate $P(\Delta s_{\mathrm{m}} \geq 20) \approx 1335/57448 \approx 2.3 \times 10^{-2}$,
which compares well with the corresponding ratio 
$P(\Delta s_{\mathrm{m}} \geq 20) \approx 71588/3316727 \approx 2.2 \times 10^{-2}$
obtained from the
direct Langevin simulations.
Also for all paths with $\Delta s_{\mathrm{m}} \geq 20$, 
we calculate the mean path from both the
 MCMC sample paths and the direct Langevin trajectories.
We show the resulting mean paths in Fig.~\ref{fig:MCMC} (b) as dash-dotted and solid orange curves,
and again observe good agreement. 
As the figure shows, paths that dissipate a lot of energy into the heat bath move
along the shear force $\Fvec$ most of the time:
The mean path first moves towards negative $x_1$ values
in the lower half-plane, and then moves towards $\xvecf$ 
in the upper half-plane.

Overall, Fig.~\ref{fig:MCMC} 
 shows that the tubular-ensemble approach to the entropy production
enables us to quantify and analyze the irreversibility of a path ensemble using only
directly measured sojourn probabilities.

\section{Discussion} 
\label{sec:discussion}
We have shown that the path-wise medium entropy
production can be obtained from the ratio of probabilities for trajectories to
remain within a tube encircling a path, in the limit of the tube radius going
to zero. As we demonstrate by analyzing  ensembles of short simulated trajectories,
using this definition the entropy production along an individual path
can be inferred from data without any knowledge of the
underlying dynamics other than assuming it to be memoryless.
By combining this measurement of irreversibility along individual paths with a Markov Chain Monte Carlo (MCMC)
algorithm, we obtain the irreversibility for  
path ensembles, using only measured sojourn probabilities.
The path ensemble we generate via our MCMC algorithm allows us to probe
the mean behavior of small- and large-entropy paths.

Our work shows clearly
that by considering individual paths as limits of finite-radius tubes,
which from an experimental point of view is a natural perspective,
both single-trajectory and path-ensemble properties
can be inferred from recorded time series without the need to parametrize a model.
Our definition of the medium entropy production, Eq.~\eqref{eq:sm_sojourn}, does not
involve non-differentiable stochastic trajectories and thus generalizes
to processes with configuration-dependent diffusivities in a manner
that side-steps delicate issues of stochastic integration (i.e.~the
It\^{o}-Stratonovic dilemma) \cite{van_kampen_ito_1981,kampen_stochastic_2007,gardiner_stochastic_2009}.
The exit rate provides information beyond the entropy production, as
Eq.~\eqref{eq:rate_and_work_rate}, with the differential $\dot{\trajvec}_{t}\mathrm{{d}}t$
chosen along $N$ linearly independent directions, can be used to measure
the drift $\Fvec$ of the process, without the need
to estimate the diffusivity. 
Our work raises the question of how the
medium entropy production could be generalized to tubes of finite
radius and what the relationship of such a definition would be to
the single-trajectory and full-ensemble measures of entropy production.
Finally, our work suggests a generalization to stochastic field theories
with broken detailed balance that are used to describe the fluctuating
dynamics of active matter \cite{cates_active_2019}.

\acknowledgements
\textbf{Acknowledgements.}
Work was funded in part by the European Research Council under the EU's Horizon 2020 Program, Grant No.~740269, and in part by the Royal Society through grant RP17002.

\appendix

\section{Sample data preparation}
\label{sec:sample_data_preparation}

We fix a length scale $L$ and a time scale $\tau$, 
and
partition $\mathbb{R}^2$ into a regular grid 
\begin{align}
\label{eq:grid}
	S_{ij} &\equiv 
	[(i-1/2 )  \Delta x,(i+1/2 )  \Delta x] 
	\\ & \qquad 
\nonumber
\times [(j-1/2 )  \Delta y,(j+1/2 )  \Delta y],
\end{align}
where $i, j \in \mathbb{Z}$.
We use $\Delta x=\Delta y = 0.05 L$,
and consider the range $-100  \leq i \leq 90$, $-60 \leq j \leq 70$.
The corresponding $S_{ij}$ then cover the domain $[-5.05L,4.5L]\times [-3L,3.5L] \subset \mathbb{R}^2$,
which is sufficient for our purposes, as we do not consider any tubes 
that extend outside this domain.

For each tuple $(i,j)$, we run 1500 independent simulations 
of the overdamped It\^{o}-Langevin Eq.~\eqref{eq:Langevin}. 
Each trajectory starts in $S_{ij}$, and we draw the initial condition from a uniform distribution on $S_{ij}$.
We then run the standard Euler-Maruyama algorithm for
 a duration $\RefillTime = 0.01\tau$ using a timestep $\Delta t = 10^{-4}\tau$.
Every simulation thus consists of $K = \RefillTime/\Delta t = 10^2$ time steps.
We use the shear force Eq.~\eqref{eq:shear_force_sum}
with $L F_0/T = 5$,
and diffusivity $D = L^2/\tau$.
Using this protocol, 
we generate one independent dataset
for each value $\theta = 0$, $0.5$, $1$, $1.5$, $2$ of the dimensionless parameter
from Eq.~\eqref{eq:shear_force_sum}.

\section{Cloning algorithm for inferring sojourn probabilities and 
exit rates from recorded time series}
\label{sec:cloning_algorithm}

We now explain how we extract finite-radius sojourn probabilities and
their associated instantaneous exit rates from a set of time series, such as
the one described in App.~\ref{sec:sample_data_preparation}.
The algorithm described here is the two-dimensional
generalization of a similar algorithm previously used 
on one-dimensional experimental time series \cite{gladrow_direct_2020},
and is illustrated in Fig.~\ref{fig:cloning_algorithm}.
We assume as given a path $\trajvec$, defined for time $[0,t_f]$,
and a tube radius $R$, as well as a dataset as described in App.~\ref{sec:sample_data_preparation}, 
with an associated timestep $\Delta t$ and trajectory length $K  \Delta t = \RefillTime$.

\textit{Initial distribution for the cloning algorithm.}
To begin, we identify the cell in which the trajectory starts, i.e.~we determine
the indices $(i_0,j_0)$ such that $\trajvec(0) \in S_{i_0,j_0}$.
We then randomly choose  $M_0$
 of the recorded trajectories from
 the cells $S_{i,j}$ with $i \in \{i_0-2,i_0-1,..,i_0+2\}$, $j \in \{j_0-2,j_0-1,..,j_0+2\}$;
the 
trajectories are chosen from a uniform distribution on all the trajectories that start within 
these cells, and with replacement.
This initial condition models a smeared-out delta-peak at the initial tube center $\trajvec(0)$.

\textit{Iteration step of the cloning algorithm.}
To infer the decay of the sojourn probability until the final time $t_f$,
the  iteration step described in the following is repeated $\mathcal{N} = t_f/\RefillTime$ times.
In the $l$-th repetition, 
the sojourn probability is obtained for $t \in [l \RefillTime, (l+1)\RefillTime]$.
For $l = 0$, $M_0$ sample time series have been selected as described above; for $l \geq 1$,
$M_l$ sample time series have been selected as will be described further below.

In the $l$-th iteration step,
we follow the $M_l$ sample time series for the duration $\RefillTime$, and 
keep track of how many sample time series have never left
the instantaneous
tube (i.e.~the moving circle with radius $R$ and
 center parametrized by $\trajvec$) between the initial time of the $l$-th iteration, $t_{l,0} \equiv l \RefillTime$, and each later
 instant $t_{l,k} \equiv l \RefillTime +  k \Delta t$, where $0 \leq k \leq K$.
We denote by $M_{l,k}$ the number of trajectories that have never left the tube until time $t_{l,k}$,
so that $M_{l,0} \equiv M_{l}$.
For the time interval $[l \RefillTime, (l+1) \RefillTime]$, 
we then approximate the sojourn probability as
\begin{equation}
\label{eq:sojourn_measure}
P_{R}^{\trajvec}(t_{l,k}) = 
\left(\prod_{m=0}^{l-1}\frac{M_{m,K}}{M_m}\right) \frac{M_{l,k}}{M_l}, \quad 0 \leq k \leq K,
\end{equation}
where for $l=0$ the product is defined as $1$ (the product describes the overall decay
of the sojourn probability until time $l \RefillTime$, i.e.~the sojourn probability until
the beginning of the current time interval $[l \RefillTime, (l+1)\RefillTime]$).

From the $M_{l,K}$ sample trajectories that have remained within
 the tube until time $t_{l,K} \equiv l \RefillTime + K  \Delta t = (l+1)  \RefillTime$,
we construct a normalized histogram using the bins $S_{ij}$ defined in Eq.~\eqref{eq:grid}.
Using this histogram as probability distribution on the cells $S_{ij}$, 
and employing a uniform distribution
for the recorded time series 
within each cell $S_{ij}$, we draw $M_{l+1}$ new time series
 from the dataset from App.~\ref{sec:sample_data_preparation}.
 In the  ($l$$+$1)-th iteration of the algorithm,
 we then follow these newly drawn trajectories.

All trajectories in the algorithm are drawn with replacement; if the initial position of a drawn
 trajectory is not within the tube initially
(which can occur if a cell only partly overlaps with the instantaneous tube),
a new trajectory is drawn from the same cell $S_{ij}$ 
until the initial condition of the sample is within the instantaneous tube.
As detailed at the end of the present appendix, 
we choose the values of $M_l$ dynamically, depending on the current trend of the
sojourn probability.

\textit{Numerical calculation of exit rate from sojourn probability.}
The decay of the sojourn probability is quantified by the instantaneous
exit rate at which trajectories first leave the tube, as defined in
Eq.~\eqref{eq:exit_def}.
To calculate the exit rate numerically,
we discretize Eq.~\eqref{eq:exit_def} using the central difference scheme with the same
timestep $\Delta t$ as used for the sample data.
We then
evaluate the time-discretized expression using the
measured sojourn probability Eq.~\eqref{eq:sojourn_measure}.

\textit{Estimating the number of samples.}
The algorithm we use to measure sojourn probabilities
 from simulations
relies on repeated random sampling 
of recorded time series. 
To choose the number of drawn samples $M_l$ efficiently, 
we employ the same algorithm as used in Ref.~\cite{gladrow_direct_2020}.
More explicitly, at the beginning of the $l$-th repetition ($l>1$) of the cloning
algorithm, we fit a linear function 
\begin{align}
\alpha_{\mathrm{fit}}(t) & =a (t-l\RefillTime)+b,\label{eq:aexit_fit}
\end{align}
to the measured exit rate in the time interval $[l\RefillTime-\Delta t_{\mathrm{fit}},l \RefillTime]$,
where $\Delta t_{\mathrm{fit}}/\td=\min\{\,0.05,\RefillTime/\td\}$.
This fit quantifies the trend of the sojourn probability in the recent past.
We use the fitted exit rate to estimate the expected decay of
the sojourn probability for the next iteration duration $\RefillTime$,
and choose $M_{l}$ such that at the end of the iteration step
we expect to have $N_{\mathrm{final}}$ trajectories remaining inside
the tube. This leads to 
\begin{align}
N_{\mathrm{final}} & =M_l~\exp\left[-\int_{l \RefillTime}^{(l+1) \RefillTime}~\alpha_{\mathrm{fit}}(s)\mathrm{d}s\right],\\
\Longleftrightarrow\qquad M_l & =N_{\mathrm{final}}~\exp\left[a\frac{\RefillTime^{2}}{2}+b\,\RefillTime\right].
\end{align}
Unless noted otherwise, we
use
$M_0 = 5 \times 10^4$ and
 $N_{\mathrm{final}}=2 \times 10^{4}$
for all data shown in the present work.

\section{Sojourn-probability MCMC algorithm for medium entropy production}
\label{sec:MCMC}

We now summarize the Metropolis-Hastings algorithm \cite{thijssen_computational_2007}
which we use for our Markov Chain Monte Carlo (MCMC) 
sampling of the transition path ensemble.
We approximate the space of transition paths from $\xveci = (0,0)$
to $\xvecf = (L,L)$ by the parametrization Eq.~\eqref{eq:path_parametrization}.
We consider $M= 15$ two-dimensional mode vectors, 
so that we run the MCMC algorithm on a space of
dimension $d=30$.

\textit{Initialization.}
As initial condition for the
 MCMC
 algorithm, we draw a random state $\avecunderline^{(0)} = \kappa \etavec$,
with $\etavec$ a sample from a $d$-dimensional normal distribution
with vanishing mean and unit covariance matrix,
\textcolor{black}{and $\kappa = 1/10$ a scaling factor that determines the covariance of the initial state
$\avecunderline^{(0)}$; we comment on our choice for $\kappa$ at the end of the present appendix.}

\textit{Monte Carlo step.}
In the $k$-th MCMC step, a candidate $\avecunderline'$ for the subsequent state $\avecunderline^{(k+1)}$ 
is proposed
 from the current state $\avecunderline^{(k)}$ via
$\avecunderline' = \avecunderline^{(k)} + \kappa \etavec, $
where $\etavec$ is drawn from a $d$-dimensional normal distribution
with vanishing mean and unit covariance matrix,
\textcolor{black}{and we use the same scaling factor $\kappa = 1/10$ as for the initialization.}
We subsequently evaluate 
the log-ratio of path probabilities for 
the paths corresponding to $\avecunderline'$, $\avecunderline^{(k)}$,
\begin{equation}
\xi \equiv
\ln \frac{P(\avecunderline')}{P(\avecunderline^{(k)}) }
\equiv
  \lim_{R\rightarrow 0} 
  \ln \frac{
  P_R^{\trajvec(\avecunderline')}
  }{
  P_R^{\trajvec(\avecunderline^{(k)})}
  },
\end{equation}
by extrapolating the log-ratio of measured
 finite-radius sojourn probabilities to the limit $R = 0$ \cite{gladrow_direct_2020}.
More explictly, 
for the paths corresponding to $\avecunderline'$, $\avecunderline^{(k)}$,
we use the cloning algorithm from App.~\ref{sec:cloning_algorithm} to 
measure the finite-radius sojourn probabilities for tube radius
 $R/L = 0.3, 0.5, 0.7$.
 This yields three datapoints for the finite-radius log-ratio
 $\ln  P_R^{\trajvec(\avecunderline')}/P_R^{\trajvec(\avecunderline^{(k)})}$,
which we extrapolate to zero by
 fitting $f(R) = a + R^2 b$
 and using 
 $\lim_{R \rightarrow 0} \ln  P_R^{\trajvec(\avecunderline')}/P_R^{\trajvec(\avecunderline^{(k)})} \equiv a$
   \cite{gladrow_direct_2020}.
To determine whether the proposed state
 $\mathbf{a'}$ is accepted, 
 we draw a random number $u$ from a uniform distribution on
 $[0,1]$.
 If
\textcolor{black}{$u \leq e^{\xi}\equiv P(\avecunderline')/P(\avecunderline^{(k)})$},
  we set $ \avecunderline^{(k+1)} = \avecunderline'$ as the next MCMC state; otherwise, 
 we use $\avecunderline^{(k+1)} = \avecunderline^{(k)}$ \cite{thijssen_computational_2007}.

\textit{Evaluation of medium entropy production.}
The MCMC algorithm yields a sequence of transition paths 
parametrized by their expansion coefficients, i.e.~$(\avecunderline^{(0)}, \avecunderline^{(1)}, ...)$.
For every 5th path we evaluate
the medium entropy production via Eq.~\eqref{eq:sm_sojourn}.
Since the forward path sojourn probabilities
 for radius $R/L = 0.3, 0.5, 0.7$
have already been measured for the MCMC step,
we only need to evaluate the corresponding backward path sojourn probabilities
to obtain the entropy production along the path.
For every 5th path 
we therefore measure the backward-path sojourn 
probabilities at radius $R/L = 0.3, 0.5, 0.7$,
 then fit a quadratic function $f(R) = a + R^2b$ to the log-ratio
$\ln P_{R}^{\trajvec(\avecunderline^{(k)})}(\tf) /{P_{R}^{\trajbackvec(\avecunderline^{(k)})}(\tf)}$,
and extrapolate
to $R \rightarrow 0$
as
$\lim_{R \rightarrow 0} 
\ln P_{R}^{\trajvec(\avecunderline^{(k)})}(\tf) /{P_{R}^{\trajbackvec(\avecunderline^{(k)})}(\tf)} 
\equiv a$.

\textit{Numerical parameters for sojourn probabilities.}
For the evaluation of all finite-radius sojourn probabilities in the MCMC algorithm
we use the 
time series from App.~\ref{sec:sample_data_preparation}
with $\theta = 1$,
and the 
algorithm from App.~\ref{sec:cloning_algorithm}
with $M_0 = 10^4$ and $N_{\mathrm{final}} = 5\times 10^3$.

To decrease the influence of initial conditions in the measurement of the sojourn probability, 
as observed at the far ends of Fig.~\ref{fig:shear_example} (c),
we do not use the delta-peak initial conditions described in
App.~\ref{sec:cloning_algorithm}.
Instead, before starting the MCMC algorithm, we for each radius $R/L = 0.3, 0.5, 0.7$, and each
of the points $\xveci$, $\xvecf$,
 consider a constant path, i.e.~we consider circles of radius $R$ around both the initial  and final points.
We use the algorithm from
App.~\ref{sec:cloning_algorithm} 
to let the delta-peak initial condition relax to the 
respective steady-state absorbing-boundary decay on those circles around $\xveci$, $\xvecf$.
We then use the corresponding normalized spatial distributions
 as initial conditions for each evaluation of the forward/backward sojourn
probability in the MCMC algorithm.

To increase the number of samples, and to decrease correlations among the samples,
we run 90 independent MCMC algorithms in parallel. 
We discard the first 1000 steps of each MCMC run to account for the
 fact that the initial condition $\avecunderline^{(0)}$
might correspond to a very atypical transition path.
After subtracting the first 1000 steps, the MCMC data comprises
287240 MCMC steps (the number of steps
 in the individual MCMC runs ranges, after subtracting the first 1000 steps,
  from 1245 to 6295).
Since we only use every 5th MCMC step to calculate a sample for the 
medium entropy production, our MCMC data in total yields
287240/5 = 57448 samples for $\Delta s_{\mathrm{m}}$.

\textcolor{black}{
We now briefly discuss our choice of the step size parameter $\kappa$.
Preliminary MCMC runs showed that, for our model system and parameters, 
in the subdomain of $\mathbb{R}^d$ for which
 $P(\avecunderline)$ and 
the product $\Delta s_{\mathrm{m}} P(\avecunderline)$
are non-negligible,
the components of the vector $\avecunderline \in \mathbb{R}^d$ are of order 1.
This means that if a typical MCMC step changes any component of the vector
$\avecunderline$ by a number much larger than 1, the algorithm will frequently try to leave the relevant
subdomain of $\mathbb{R}^d$ within a single step, which leads to a low MCMC acceptance rate,
and hence a large number of MCMC steps necessary to explore the relevant domain.
On the other hand, if a typical MCMC step changes the components of the vector
$\avecunderline$ only by a number much smaller than 1, 
it will take a lot of steps to explore the relevant domain.}

\textcolor{black}{
The above heuristic arguments motivate our
 choice  $\kappa = 1/10$ for the MCMC step:
 The factor $1/10$ means that 
  in each MCMC step we attempt  to
vary each vector component of $\avecunderline$
on a scale one order of magnitude smaller as compared to
   the relevant subdomain.
Ultimately, the justification for our choice of $\kappa$ is that our MCMC data is reasonably 
converged,
as discussed in App.~\ref{sec:MCMC_further_analysis} and
 in particular in Fig.~\ref{fig:MCMC_app} (a) below.
 }
 
 \textcolor{black}{
We have chosen $\kappa = 1/10$ also in our initial condition so as to be consistent with
our MCMC step.
Note that, 
since we disregard the first 1000 steps of each MCMC run, the details of the initial condition are in 
fact not important for our result, as long as the initial values for each component of $\avecunderline^{(0)}$ are of the order of unity.
}

\section{Further analysis of the deviation between the entropy distributions obtained 
from sojourn-probability MCMC 
and direct Langevin simulations}
\label{sec:MCMC_further_analysis}

\begin{figure*}[ht!]
\centering \includegraphics[width=\textwidth]{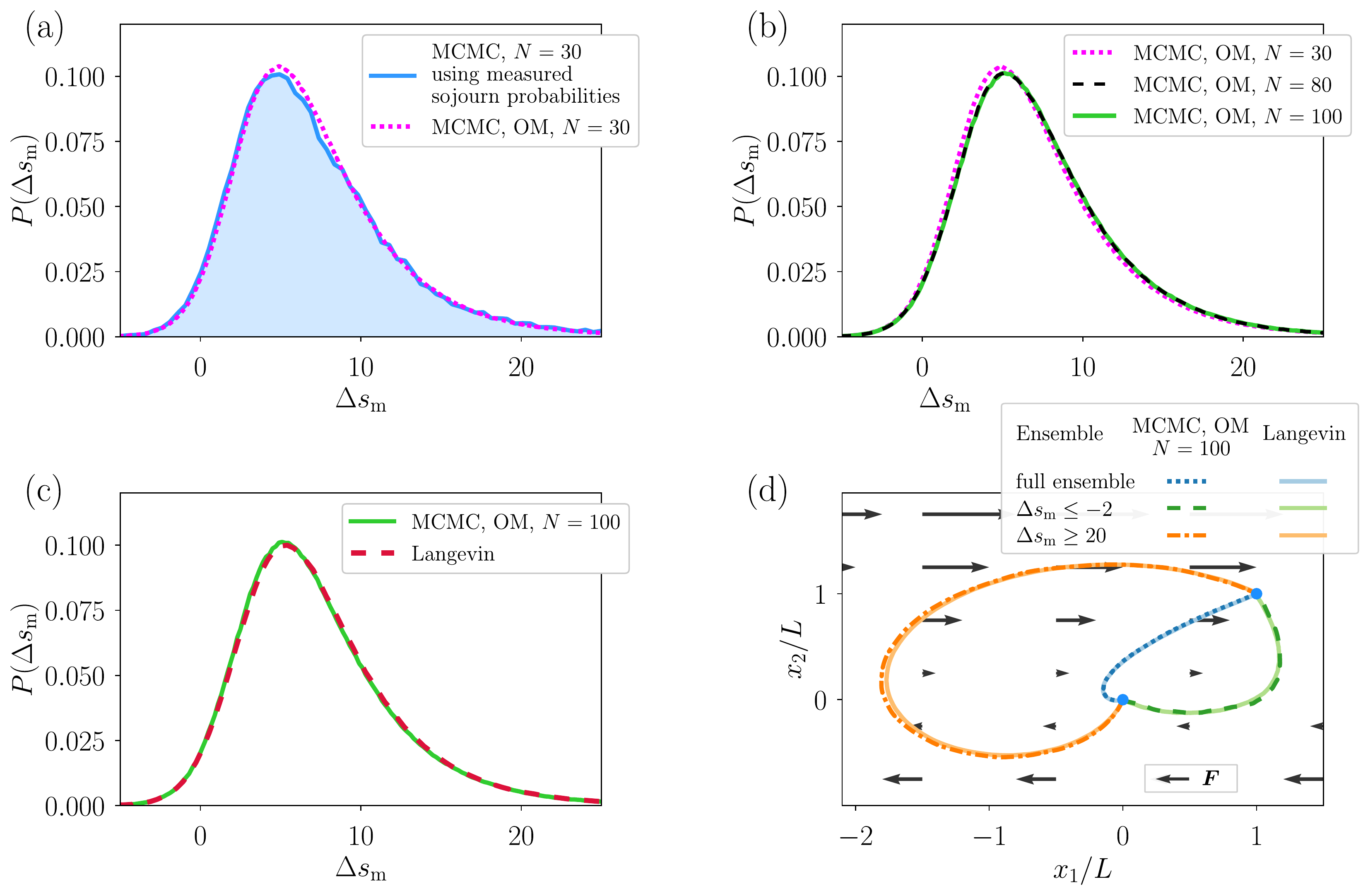}
\caption{
\label{fig:MCMC_app} 
\textit{(a)} 
The solid blue line is a replot of the 
 medium-entropy-production distribution shown in Fig.~\ref{fig:MCMC} (a),
 which is inferred using a MCMC algorithm that uses measured sojourn probabilities,
 c.f.~App.~\ref{sec:MCMC}.
The dotted magenta line displays the  medium-entropy-production distribution obtained from 
using the same MCMC algorithm,
but evaluating ratios of path probabilities and the medium entropy production
using analytical expressions, c.f.~App.~\ref{sec:MCMC_further_analysis}; consequently, 
the ``OM'' in the legend
refers to the Onsager-Machlup action Eq.~\eqref{eq:OnsagerMachlup}.
Both algorithms use the path-space parametrization Eq.~\eqref{eq:path_parametrization} with $M=15$.
\textit{(b)}
The distribution of the medium entropy production for the ensemble of paths
that start at $\xveci = (0,0)$ and at time $t = \tau_D$ are at $\xvecf = (L,L)$.
Distributions are obtained using the theoretical MCMC algorithm from App.~\ref{sec:MCMC_further_analysis},
 with $M=15$ (dotted magenta line, replot from subplot (a)),
 $M = 40$ (dashed black line), and $M=50$ (solid green line).
 \textit{(c)}
The solid green line is a replot of the $M=50$ data from subplot (c).
The dashed red line is a replot of the corresponding data from Fig.~\ref{fig:MCMC} (a),
and shows the medium entropy production for the ensemble of paths
that start at $\xveci = (0,0)$ and at time $t = \tau_D$ are at $\xvecf = (L,L)$,
obtained from evaluating the formula Eq.~\eqref{eq:theory_entropy_production_general}
 on trajectories generated from
 direct Langevin simulations.
\textit{(d)}
The solid colored lines are replots of the corresponding data
 in Fig.~\ref{fig:MCMC} (b).
The blue dotted and solid curves represent the mean path for the full ensemble of paths that move from
$\xveci = (0,0)$ to $\xvec_{f} = (1,1)$ in a time $t = \tau_D$.
While the dash-dotted and solid orange lines show the
 mean over such paths with entropy production $\Delta s_{\mathrm{m}} \geq 20$,
the dashed and solid green lines denote the mean for low-entropy production paths with $\Delta s_{\mathrm{m}} \leq -2$. 
All solid lines are obtained 
 from direct Langevin simulation,
the corresponding broken lines represent means for the $N=100$ theoretical MCMC 
data, c.f.~subplots (b), (c).
}
\end{figure*}
In Fig.~\ref{fig:MCMC} (a), we consider the distribution of the medium entropy production,
obtained i) from the sojourn-probability MCMC algorithm described in App.~\ref{sec:MCMC},
and ii) from the theoretical formula, Eq.~\eqref{eq:theory_entropy_production_general}, evaluated on directly simulated 
Langevin trajectories.
In the present appendix, we demonstrate that the 
slight differences in the two datasets can be explained
by the number of modes used in the parametrization Eq.~\eqref{eq:path_parametrization},
which, for the MCMC data shown in Fig.~\ref{fig:MCMC}, is $M=15$.

For this, 
we consider a variation of the MCMC algorithm described in App.~\ref{sec:MCMC}.
In this variation, we do not use recorded sample trajectories to evaluate ratios of path probabilities
and the entropy production, but instead use 
i) the difference in Onsager-Machlup actions Eq.~\eqref{eq:OnsagerMachlup} for
log-ratios of path probabilities,
and ii)
the analytical formula, 
 Eq.~\eqref{eq:theory_entropy_production_general},
for the medium entropy production.
To distinguish it from the data-driven MCMC algorithm described in App.~\ref{sec:MCMC},
we in the following refer to this MCMC algorithm as ``theoretical MCMC'';  here 
``theoretical'' means
 that neither path probabilities nor entropy productions are measured from data, but rather
 evaluated using the corresponding analytical formulas available for overdamped Langevin dynamics.
We run the theoretical MCMC 
using the same parameters for the cloning algorithm as in App.~\ref{sec:MCMC}.
For each parameter combination considered below, 
we run 100 independent theoretical MCMC realizations with 200000 steps each, and
evaluate the medium entropy production for every MCMC step.
We discard the first 999 MCMC steps for each run, which means that 
for each parameter combination, our theoretical MCMC ensemble
 consists of in total $100 \times (199001) \approx 1.99\times 10^7$ datapoints for $\Delta s_{\mathrm{m}}$.
 Thus, for the theoretical MCMC we have two orders of magnitude more
MCMC paths as compared to the data-driven MCMC results shown in Fig.~\ref{fig:MCMC}, which is because
the theoretical MCMC is computationally much cheaper.

We now show that the deviations between the two curves in Fig.~\ref{fig:MCMC} (a)
originate from the relatively low number of modes we use, $M =15$.
For this, we consider the theoretical MCMC 
with also $M=15$ modes, i.e.~the same number of modes as used for the sojourn-probability
MCMC in Fig.~\ref{fig:MCMC} (a).
We compare the sojourn-probability- and theoretical-MCMC results
in Fig.~\ref{fig:MCMC_app} (a), where we observe that the distributions are
very similar, with only minor deviations around $\Delta s_{\mathrm{m}} \approx 5$.
This indicates that the sojourn-probability MCMC data is sufficiently converged,
and that the deviations from the direct Langevin simulations are 
due to the low-dimensional approximation $M=15$ of the path space.
We chose $M=15$ in the main text
 as a compromise between approximation error (which decreases
with increasing $M$) and
convergence speed of the MCMC algorithm (which decreases with increasing $M$).

We additionally run the theoretical MCMC algorithm
for $M=40$ and $M=50$ mode vectors, corresponding to $N=M  d = 80$, $100$, respectively.
The resulting distributions
are shown in Fig.~\ref{fig:MCMC_app} (b), where we observe that the $M=15$ data slightly disagrees
with the $M=40$, $50$ results.
This confirms that the projection on only $M=15$ modes leads to a 
distortion of the actual distribution of the medium entropy production.
The distributions for $M=40$ and $M=50$ modes agree with each other very well,
so that we conclude that $M \geq 40$ modes
 are enough to reproduce the actual distribution.
Indeed, the $M=50$ theoretical MCMC data agrees very well with the direct Langevin results,
see Fig.~\ref{fig:MCMC_app} (c).

In Fig.~\ref{fig:MCMC_app} (d), we finally compare the mean paths obtained from direct Langevin simulations,
and shown in Fig.~\ref{fig:MCMC} (b),
to the corresponding mean paths of the $M=50$  theoretical MCMC data.
We observe that all three path pairs are in very good agreement;
this once again confirms the validity of the MCMC algorithm.

\section{Entropy production along closed loops in a circular double well}

\label{app:double_well}

\begin{figure*}[ht!]
\centering \includegraphics[width=0.45\textwidth]{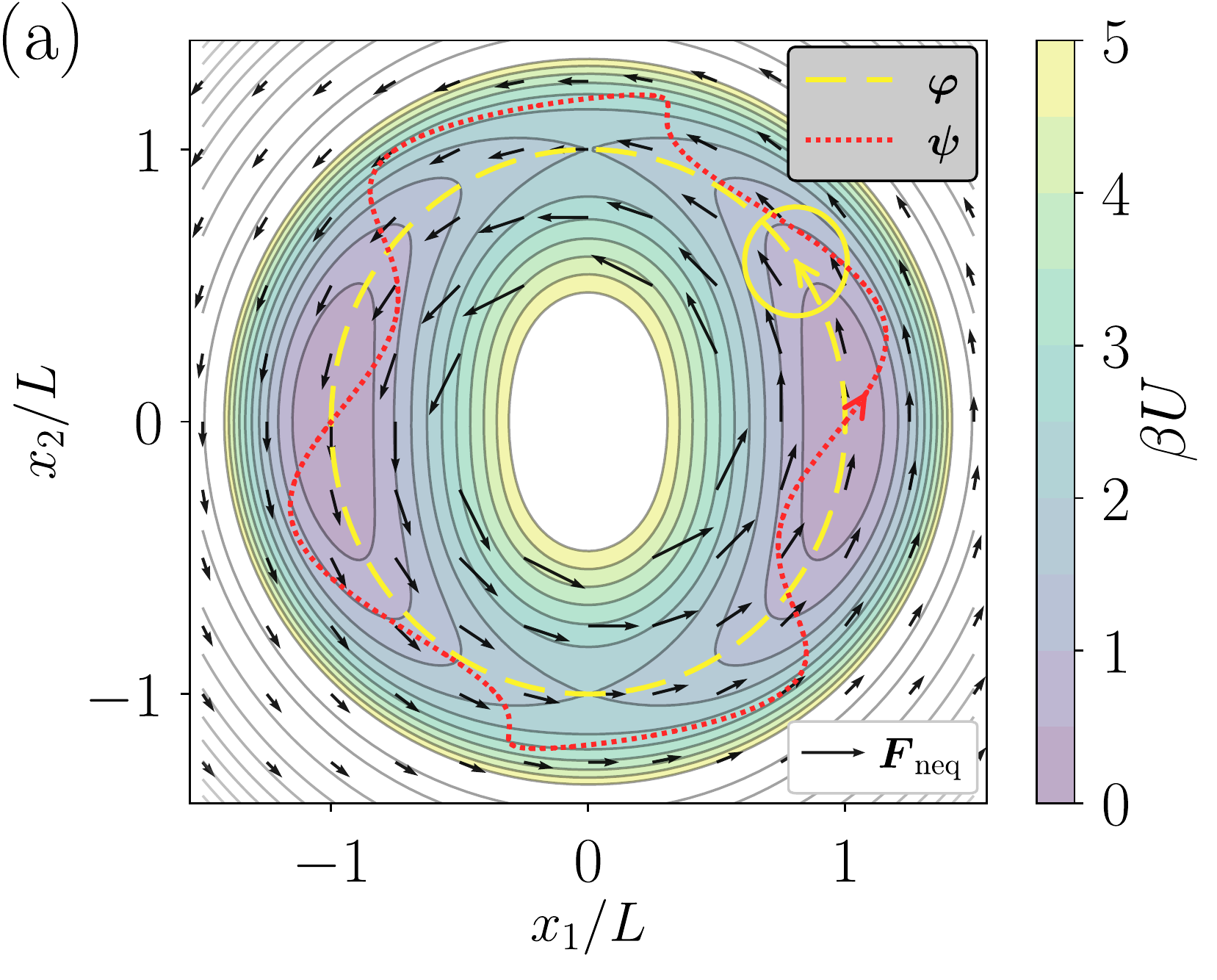}
\qquad{}\includegraphics[width=0.43\textwidth]{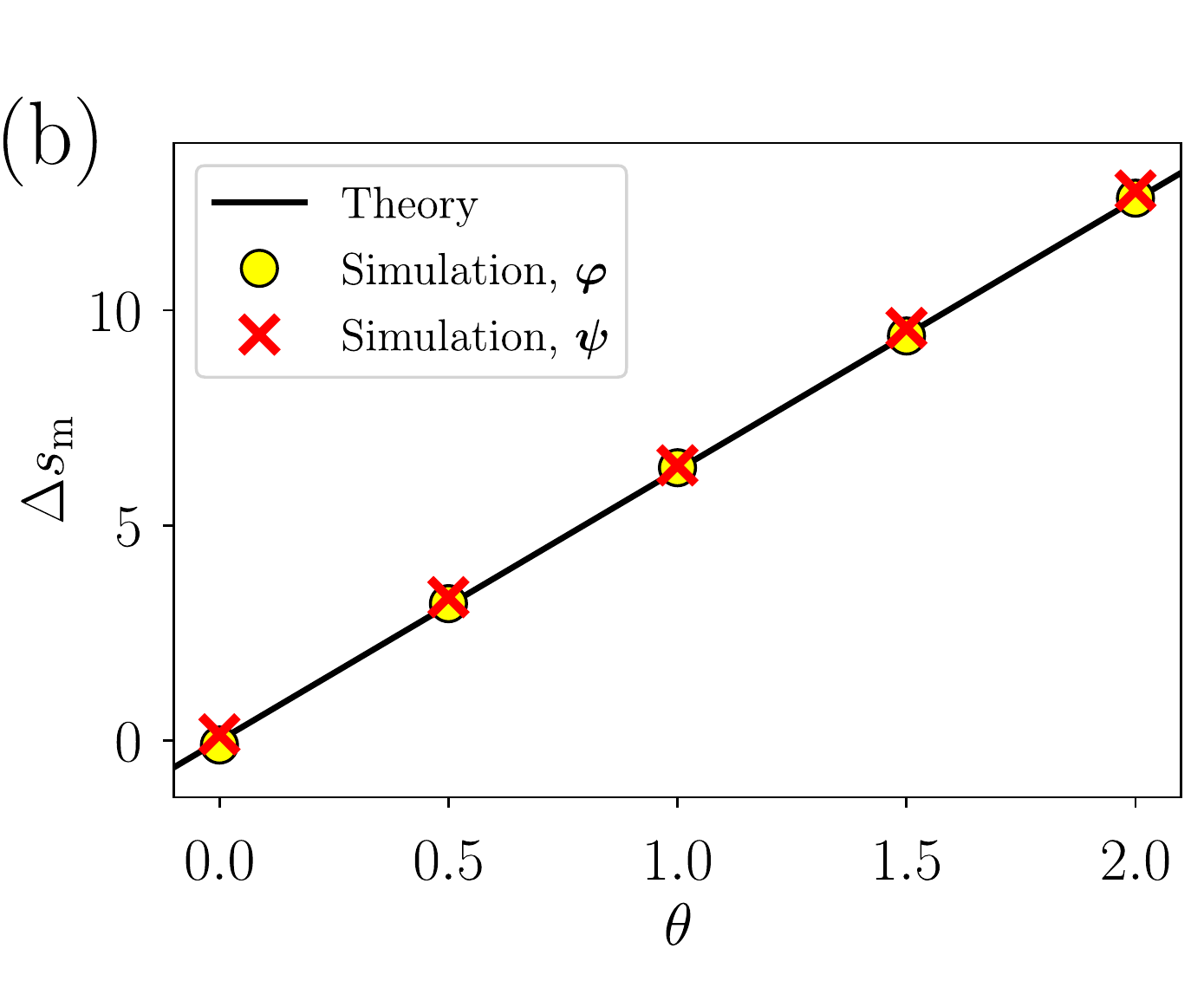}
\\
 \includegraphics[width=0.45\textwidth]{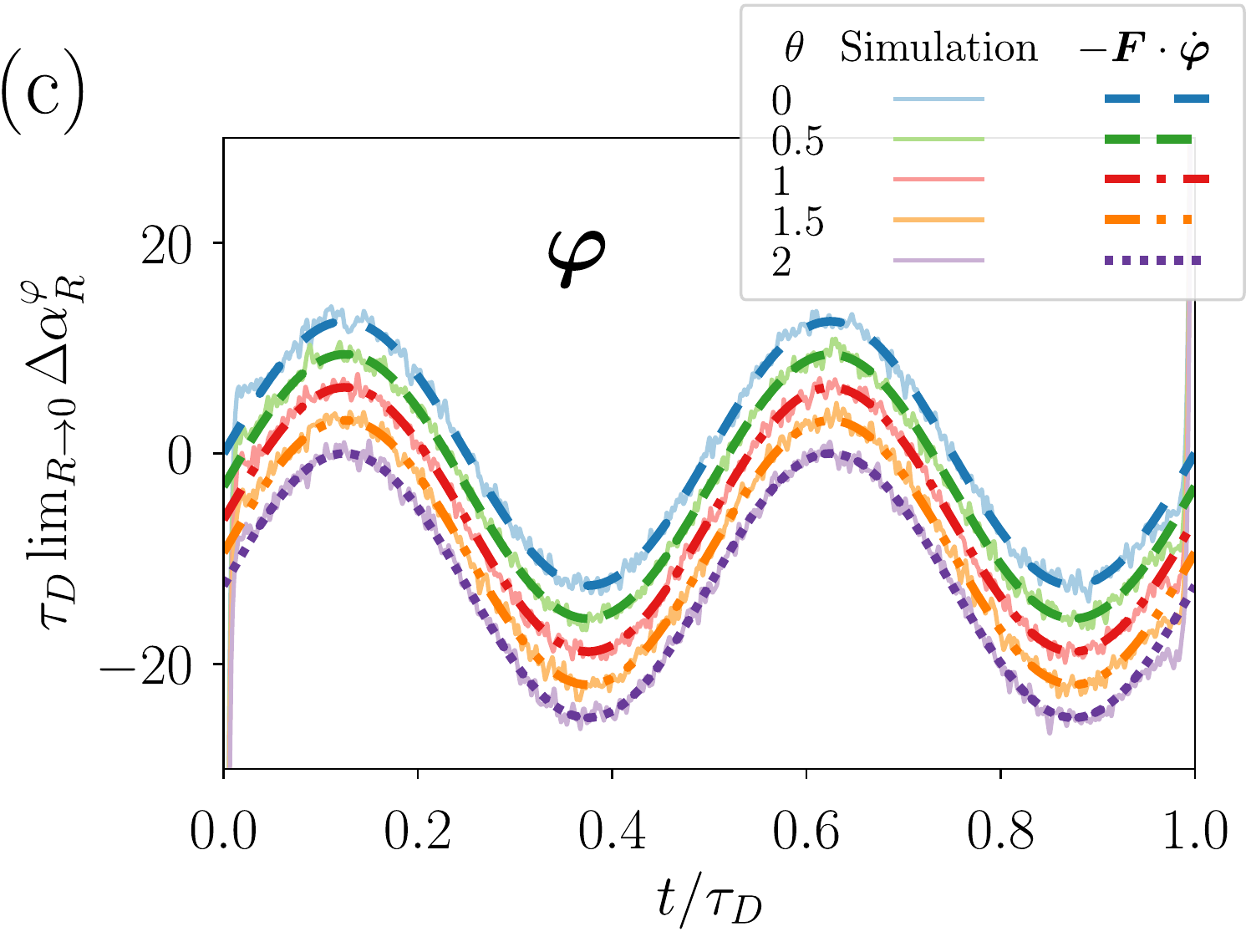} \qquad{}\includegraphics[width=0.45\textwidth]{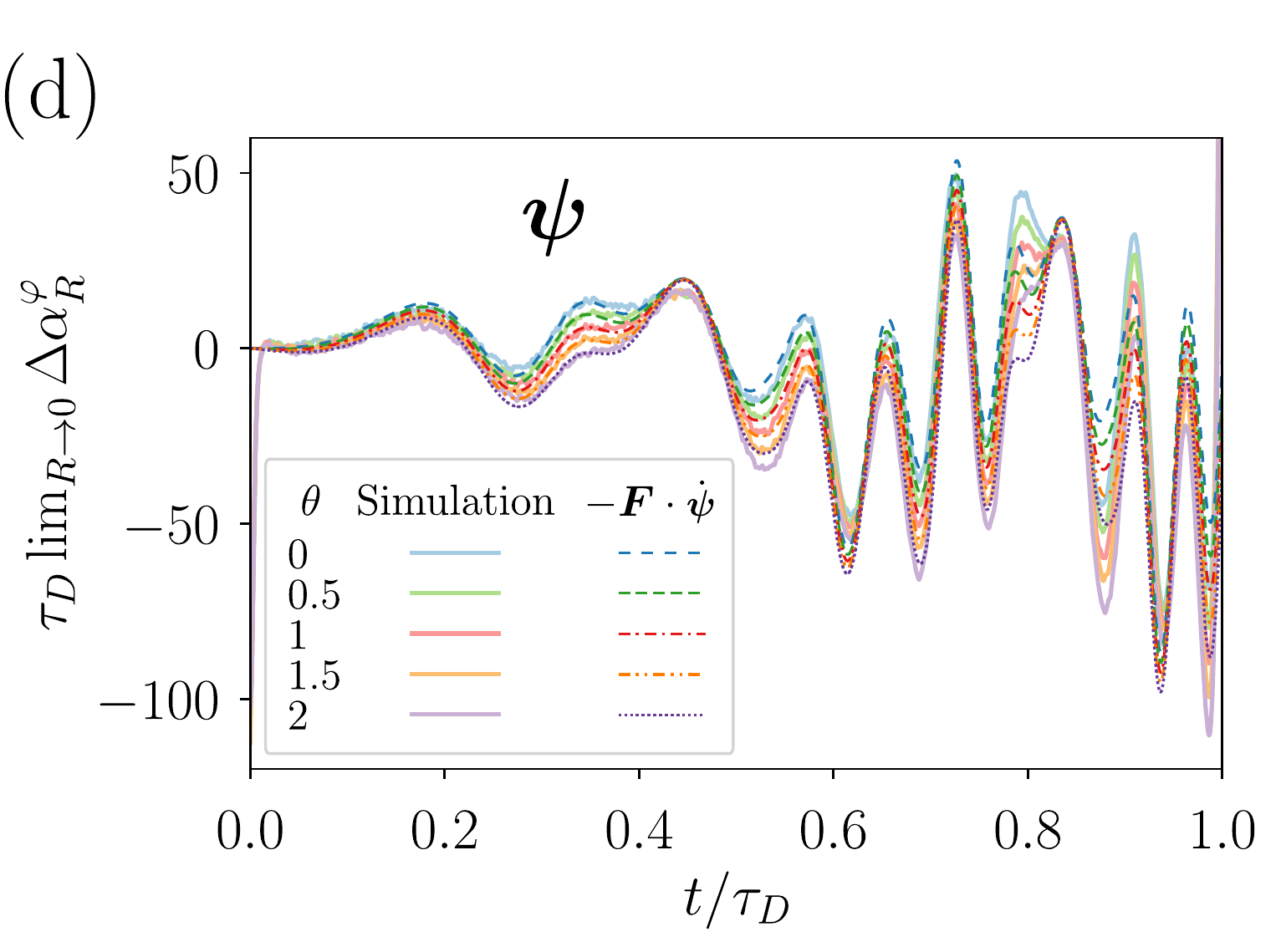}
\caption{\label{fig:circle} \textit{(a)} The colored contours show the potential
$U$ defined in Eq.~\eqref{eq:U_def}. The non-equilibrium force
Eq.~\eqref{eq:Fneq} is shown as black quiver plot. The dashed yellow 
line denotes the path $\trajvec$ defined in Eq.~\eqref{eq:path0_def},
the yellow circle indicates an instantaneous ball of radius $R/L=0.2$
around $\trajvec$. The dotted red line denotes the path $\trajTwovec$
defined in Eq.~\eqref{eq:path1_def}. For both paths $\trajvec$,
$\trajTwovec$, arrows indicate the forward direction. \textit{(b)}
The black line denotes the theoretical entropy production Eq.~\eqref{eq:theory_entropy_production}
for $\Gamma=1$. The colored symbols denote the entropy production
obtained by evaluating the right-hand side of Eq.~\eqref{eq:sm_sojourn}
using extrapolated measured sojourn probabilities.
The dots correspond to $\trajvec$, the crosses are obtained
using $\trajTwovec$. \textit{(c), (d)} The solid colored lines denote
the extrapolation to $R\rightarrow0$ of measured finite-radius exit-rate
differences between forward- and backward paths, for several values
of $\theta$ and the reference path (c) $\trajvec$ and (d) $\trajTwovec$.
The broken colored lines denote the corresponding theoretical predictions
given by the right-hand side of Eq.~\eqref{eq:theory_entropy_production_general},
calculated using the force Eq.~\eqref{eq:force_sum}. Numerical data
is smoothed using a Hann window of width $0.005\,\td$. 
}
\end{figure*}
We here consider a second example system. For a length scale $L$
and a time scale $\timescale$, we again consider the overdamped It\^{o}-Langevin
Eq.~\eqref{eq:Langevin} for dimension $N=2$ with diffusivity $D=L^{2}/T$,
so that $\tau_{D}\equiv L^{2}/D=\timescale$. We now consider a force
$\Fvec$ 
\begin{equation}
\Fvec(\xvec)=-\left(\nablavec U\right)(\xvec)+\theta\,\Fvec_{\mathrm{neq}}(\xvec),\label{eq:force_sum}
\end{equation}
which is given as a sum of the gradient of a potential $U$ and an
additional term $\Fvec_{\mathrm{neq}}$ which is non-conservative,
i.e.~does not admit a (global) potential. As in the main text, the
dimensionless parameter $\theta\in\mathbb{R}$ controls the amplitude
of the non-conservative force, and for $\theta\neq0$ this system
is a non-equilibrium system. For $U$ we consider a sombrero potential
superimposed with an angular double well, defined as 
\begin{align}
U(\xvec) & =U_{0}\left[\left(\frac{||\xvec||}{L}\right)^{2}-1\right]^{2}\label{eq:U_def}\\
 & \quad+U_{1}\left[\frac{1+\cos(2\phi)}{2}\dfrac{1+\exp\left(\dfrac{1}{10}\right)}{1+\exp\left(\dfrac{||\xvec||^{2}}{10L^{2}}\right)}-1\right],\nonumber 
\end{align}
where $||\xvec||=\sqrt{x_{1}^{2}+x_{2}^{2}}$, $x_{1}=||\xvec||\cos(\phi)$,
$x_{2}=||\xvec||\sin(\phi)$; we use $\beta U_{0}=5$, $\beta U_{1}=2$.
This potential, which is illustrated in Fig.~\ref{fig:circle} (a),
has local minima at $\xvec=(L,0)$, $(-L,0)$, and saddle points at
$\xvec=(0,L)$, $(0,-L)$. For the non-equilibrium force we consider
an angular force 
\begin{equation}
\frac{1}{\kT}\Fvec_{\mathrm{neq}}(\xvec)=\frac{1}{||\xvec||^{2}}\begin{pmatrix}-x_{2}\\
~~x_{1}
\end{pmatrix},\label{eq:Fneq}
\end{equation}
which illustrated as a quiver plot in Fig.~\ref{fig:circle} (a).

For the force Eqs.~\eqref{eq:force_sum}, \eqref{eq:U_def}, \eqref{eq:Fneq},
and $\trajvec$ a closed loop, the analytical entropy production Eq.~\eqref{eq:shear_force_sum}
is given by 
\begin{equation}
\Delta s_{\mathrm{m}}[\trajvec]=2\pi\Gamma\theta,\label{eq:theory_entropy_production}
\end{equation}
where $\Gamma\in\mathbb{Z}$ is the winding number which quantifies
how often the path $\trajvec$ winds counterclockwise around the origin
$\xvec=\boldsymbol{0}$. Thus, for the particular nonequilibrium force
Eq.~\eqref{eq:Fneq}, the theoretical entropy production Eq.~\eqref{eq:theory_entropy_production}
is topological, i.e.~only depends on the winding number and not on
more details of the path.

We consider two circular paths 
\begin{align}
\trajvec_{t} & =L\begin{pmatrix}\cos\left(2\pi t/\tfinal\right)\\[1.2ex]
\sin\left(2\pi t/\tfinal\right)
\end{pmatrix},\label{eq:path0_def}\\
\trajTwovec_{t} & =L\begin{pmatrix}\cos\left(2\pi t^{2}/\tfinal^{2}\right)\\[1.2ex]
\sin\left(2\pi t^{2}/\tfinal^{2}\right)
\end{pmatrix}+\frac{L}{5}\begin{pmatrix}\sin\left(10\pi t^{2}/\tfinal^{2}\right)\\[1.2ex]
\sin\left(2\pi t^{2}/\tfinal^{2}\right)
\end{pmatrix},\label{eq:path1_def}
\end{align}
where $t\in[\ti,\tf]\equiv[0,\timescale]$. These paths, which both
have a winding number $\Gamma=1$, are shown in Fig.~\ref{fig:circle}
(a) as yellow dashed and dotted red lines.

For $\theta=0,0.5,1,1.5,2$ and $R/L=0.2,0.25,0.3,0.35,0.4,0.45,0.5$,
we measure the entropy production along the forward- and reverse version
of each path $\trajvec$, $\trajTwovec$, using 
a variation of the cloning algorithm
from App.~\ref{sec:cloning_algorithm}:
Instead of binning space, and creating a set of sample 
time series beforehand,
we run simulations on the fly.
Initial conditions for the $(n+1)$-th iteration are then sampled from a uniform distribution 
on the final positions
of the trajectories that have never left the tube in the $n$-th iteration.
Also here, simulations are run using the standard Euler-Maruyama, but since we now
consider smaller tube radii, we also use a
smaller timestep $\Delta t/\td = 10^{-5}$, as well as shorter iteration times
$\RefillTime/\td = 0.005$. 
We furthermore use $M_0 = N_{\mathrm{final}} = 10^{5}$,
and a delta-peak initial condition at the initial tube center.

We extrapolate
the resulting measured finite-radius exit-rate differences between forward-
and reverse path $R=0$ as described in the main text, and in Fig.~\ref{fig:circle}
(c), (d) show that the result agrees well with the theoretical prediction
Eq.~\eqref{eq:theory_entropy_production_general} along the paths. Finally,
 in Fig.~\ref{fig:circle}
(b) we compare the negative temporal integral of the extrapolated
exit-rate differences with the expected theoretical entropy production,
and find that the numerical and theoretical results agree very well.
Thus, also this second example confirms that 
Eqs.~\eqref{eq:sm_sojourn}, \eqref{eq:definition_path_entropy_in_terms_of_exit_rate},
can be used to infer and analyze the medium entropy production along individual
paths directly from exit rates.

\end{document}